# Unlearning-Enhanced Website Fingerprinting Attack: Against Backdoor Poisoning in Anonymous Networks

Yali Yuan     Kai Xu     Ruolin Ma     Yuchen Zhang

School of Cyber Science and Engineering, Southeast University

## ABSTRACT

Website Fingerprinting (WF) is an effective tool for regulating and governing the dark web. However, its performance can be significantly degraded by backdoor poisoning attacks in practical deployments. This paper aims to address the problem of hidden backdoor poisoning attacks faced by Website Fingerprinting attack, and designs a feasible mothed that integrates unlearning technology to realize detection of automatic poisoned points and complete removal of its destructive effects, requiring only a small number of known poisoned test points. Taking Tor onion routing as an example, our method evaluates the influence value of each training sample on these known poisoned test points as the basis for judgment. We optimize the use of influence scores to identify poisoned samples within the training dataset. Furthermore, by quantifying the difference between the contribution of model parameters on the taining data and the clean data, the target parameters are dynamically adjusted to eliminate the impact of the backdoor attacks. Experiments on public datasets under the assumptions of closed-world (CW) and open-world (OW) verify the effectiveness of the proposed method. In complex scenes containing both clean website fingerprinting features and backdoor triggers, the accuracy of the model on the poisoned dataset and the test dataset is stable at about 80%, significantly outperforming the traditional WF attack models. In addition, the proposed method achieves a 2-3 times speedup in runtime efficiency compared to baseline methods. By incorporating machine unlearning, we realize a WF attack model that exhibits enhanced resistance to backdoor poisoning and faster execution speeds in adversarial settings.

# 1. Introduction

Website Fingerprinting attack is a technique that infers the private information within anonymous communication systems by monitoring the traffic characteristics of websites, thereby enabling surveillance of the dark web. At present, state-of-the-art deep learning WF methods can automatically extract and learn traffic features. Common models include convolutional neural networks (DF[Sirinam et al.[3]]), triplet neural networks (TF[Sirinam et al[5]]), and the long Short-Term memory networks (AWF[Rimmer et al. [6]]) and so on. However, it faces many challenges in practical applications. Data poisoning from malicious attackers is a human factor overlooked by many WF attack techniques, which can lead to problems such as changes in traffic characteristics and label errors in the training samples, thereby affecting the robustness and generalization of the model. In terms of model training, these models that need to be analyzed through neural networks usually have a high computational cost during training and require long-term computations on multiple GPUs. Consequently, many researchers outsource the training process to cloud servers or rely on pre-trained models to make adjustments for specific tasks. These outsourced cloud services introduce new challenges: some saboteurs can create a maliciously trained network to carry out backdoor attacks on models that adopt deep learning [Gu et al.[36]]. These attacked models have the advanced performance on the training and validation samples of users but perform poorly on the inputs selected by specific saboteurs. Gu et al. [35] indicate that backdoors in neural networks are powerful, and because the behavior of neural networks is difficult to explain, they are not easily detected and cleared by victims. Furthermore, backdoor poisoning attacks are relatively covert. That is, they evade standard validation tests and do not introduce any structural changes to neural networks trained on clean datasets. However, they can ensure that the model outputs predictions completely different from the correct results after inputting backdoor triggers [35][36].

Data collection challenges also facilitate backdoor attacks. Given the characteristics of WF, attackers' data generally cannot be self-generated. Most of them come from network traces during dark web communication processes, such as communication requests initiated by the client to the server, feedback traffic from the server to the client, communication data packets from Tor onion routing relay nodes, etc. This provides a new way for backdoor attackers to get infected. They can deploy backdoor models on the servers of websites that users frequently visit. Whenever WF analysts collect data from these websites, the resulting training dataset may inadvertently include not only clean WF information and labels but also a part of the WF sequence containing backdoor triggers. These triggers are often unrelated to the visited website and closely associated with the target website that backdoor attackers want to obfuscate [Liang et al.[36]]. Subsequently, when WF analysts want to use the trained model to predict the behavior of users visiting the website, the backdoor models deployed on the server will randomly add triggers to the traffic data packets, thereby misleading the classifier.

Our work aims to address the above-mentioned problems and propose feasible solutions by integrating the unlearning approach. In the preparatory wor, the experimental dataset was divided into a clean training set and a poisoned training set using the backdoor poisoning attack method against website fingerprintings. The toxic training dataset is mainly reflected in the characteristic alterations of data points and label errors, that is, injecting backdoor triggers into selected training points to associate their fingerprints with unrelated website labels. After dataset processing and augmentation, the two datasets are mixed to perform the initial training of the model, resulting in a poisoned model contaminated by backdoor poisoning.

To identify poisoned data points, we first select several known poisoned test points that cause model mispredictions during testing. By using the calculation function of the influence score, we quantify the influence scores of every training data points on these poisoned test point. Then, we apply random transformations to these test points. The transformation contents mainly include: randomly matching a certain number of test points with different website labels, and performing enhancement operations such as insertion, splitting, merging, and reversal of network traces for each test traffic unit with a certain probability. After transformation, we recalculate the influence score of the training WF points on the transformed test points. If the difference in the changes of the two influence scores exceeds an experimentally determined threshold, we consider that the training sample has been injected with a backdoor trigger. This process can efficiently and accurately distinguish the poisoned dataset from the clean dataset, and perform subsequent operations respectively as the forgotten dataset and the retained dataset in the next unlearning module.

We adopt the Fisher Information Matrix[8] to quantitatively estimate the importance and sensitivity of each model parameter for the retained dataset and the target (poisoned) dataset respectively. Parameters exceeding a chosen threshold are selected for selective adjustment. The threshold allows fine-grained control of the trade-off between performance and efficiency in the experiment. The specific adjustment method is reflected in suppressing the Top-K parameter values that exhibit high sensitivity to the target (poisoned) dataset but relatively low sensitivity to the retained (clean) dataset. Finally, the adjusted parameters of WF model are returned. In the unlearning process, the use of the Fisher Information Matrix can achieve efficient estimation of parameter importance, significantly reducing or eliminating the need for full model retraining and saving substantial computational time.

Our main contributions are as follows:

**1. Apply unlearning technology in the field of WF:** It is proposed to intergrate machine unlearning with Website Fingerprinting, enabling the WF classifiers to maintain functionality amidst adversarial settings characterized by malicious saboteurs and backdoor injection attacks. While current state-of-the-art WF attacks leverage deep learning to handle WF dynamics, they critically overlook the vulnerability of model training data to human-orchestrated poisoning threats. Applying unlearning technology to the field of WF attack is an effective and novel solution to this fundamental security challenge within the WF domain.

**2. Few-shot poisoned point detection method:** few-shot detection can establish quantitative connections between a small number of poisoned test points and the training dataset, and complete the extraction of poisoned data points without the need for manual inspection of the entire WF dataset. It greatly improves the detection efficiency of backdoor poisoning, making our solution far more practical and deployable in real-world scenarios.

**3. Higher backdoor poisoning and forgetting capabilities:** Our method achieves superior capabilities in both removing backdoor poisoning and eradicating its residual harmful effects. After being injected by a backdoor, there are some harmful attributes in the data that affect the judgment of unknown classes. We directly mitigate the influence of poisoned points on the classifier by strategically adjusting core model parameters, thereby ensuring a more complete and robust elimination of the backdoor's impact compared to conventional approaches.

**4. More efficient WF model:** Not only in the unlearning part but also throughout the entire WF analysis framework, our method achieves less space and time consumption compared to traditional algorithms while maintaining or even improving performance.

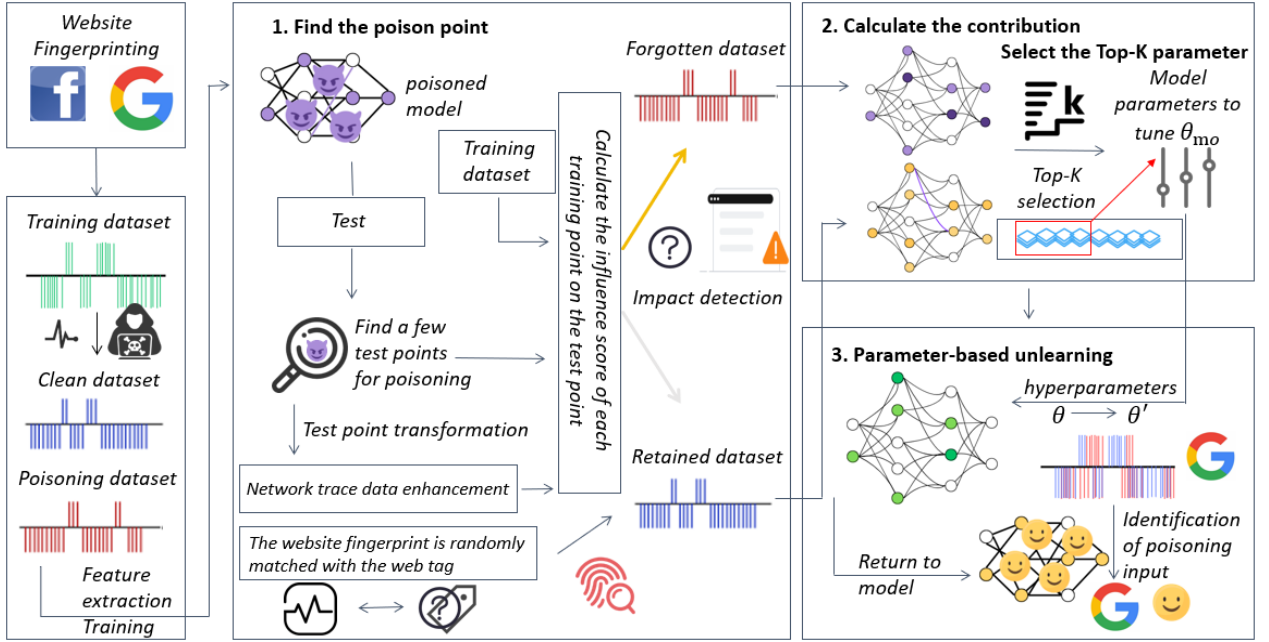

Figure 1: Our model. In the early stage, the training dataset is divided into clean and poisoned subsets by the backdoor attack method designed by us for WF. The model then leverages the resulting poisoned model to detect poisoned data points within the training set. Based on this detection, the dataset is segmented into retained (clean) and forgotten (poisoned) subsets. Subsequently, we calculate the contribution between data points and parameters is calculated, enabling the selection of target parameters for strategic adjustment. Finally, the WF classifier that can effectively eliminate and resist backdoor triggers is obtained.

Our experiments show that our method outperforms the existing algorithms of the Website Fingerprinting attack in the backdoor poisoning scenario, and can maintain a high classification accuracy on the clean dataset while resisting the poisoning effect brought by the backdoor attacks.

## 2. Preliminaries

Our model and scheme will be carried out for the unlearning process and Website Fingerprinting attacks. The following are some of the symbols and parameters that will be used later: $D_{tr} = \{x_i, y_i\}_{i=1}^{N}$ denote the entire training dataset, each of these training samples $x_i \in \{x_j\}_{j=1}^{N}$, consists of $n$ Tor units arranged in order, defined as $x_i = \{t_j\}_{j=1}^{n}$, and they all have a label $y_i \in \{y_j\}_{j=1}^{M}$ associated with them, $N$ represents the training dataset size, $M$ denote the number of tag types, which is the number of websites from which WF analysts can collect traffic in advance, also the size of the target set of websites from which dark web users can be monitored.

$D_{va} = \{x_i, y_i\}_{i=1}^{N_V}$, $D_{te} = \{x_i, y_i\}_{i=1}^{N_t}$ denote the validation set and the test set in the WF dataset.

In the dataset preparation stage, we employ a backdoor poisoning attack to induce the training dataset $D_{tr}$ to produce the predicted label error of the final model due to feature disorder [35]. The poisoning attack is defined as $P = \{p_{dict(i)}\}_{i=1}^{N_P}$, $N_P$ denote the poisoned dataset size, $dict(i)$ denote the label of the $i^{th}$ poisoning data point in the initial training dataset $D_{tr}$. This part of the dataset affected by the poisoning attack is defined as $D_{po} \subset D_{tr}$, where every poisoned data point $(x_i', y_i') \in D_{po}$, $x_i' = p_{dict(i)}(x_i)$, $y_i' \in \{y_j\}_{j=1}^{M}$ and $y_i' \neq y_i$. The remained part is not affected by the poisoning attack and can be normally input to participate in the training dataset is defined as $D_{cl} = D_{tr}/D_{po}$. Define $D_{te}' = \{x_i', y_i\}_{i=1}^{N_t}$, the input samples in this dataset are all introduced with backdoor triggers by transform $P$, which can verify the performance of the WF analysis model on the erroneous dataset.

The poisoning spot detection module is used to generate the forgotten dataset, which is defined as $D_{fo} \subseteq D_{tr}$. It represents the part of the dataset that needs to be forgotten in the machine unlearning module. After removing the influence of the forgotten dataset on the model, the remained part of the dataset that needs to maintain the performance of the original model is called the retained dataset, which is defined as $D_{re} = D_{tr}/D_{fo}$. For the detection module, our goal is to make $D_{fo}$ achieve a similar model effect as training with the dataset $D_{po}$ to the greatest extent and $D_{re}$ can approach the similar model effect as training with the dataset $D_{cl}$. The former ensures that the model can accurately forget the wrong data points and improve the accuracy of recognition on the wrong dataset. The latter requires that the model can maintain the performance ability of the original WF model on the clean dataset as much as possible.

In the machine unlearning module, the goal is that the final WF classifier model can output the original correct results on $D_{te}'$ and can maintain the performance on $D_{te}$ as much as possible as the model trained with the clean dataset, in addition to being limited to the test set. The unlearning WF model can also achieve similar results on the validation set and the training set, which does not affect the recognition accuracy of the model on the dataset without errors, and can also improve the running efficiency.

The model prediction function can be defined as follows:

$$F(wf,\theta): X \to Y \tag{1}$$

where $X \in \mathbb{R}^N$ is the input space, $Y \in \mathbb{R}^M$ is the output space, $wf$ is the WF classifier model, and $\theta$ is the parameter used to parameterize the prediction function and is continuously optimized on $D_{tr}$.

Then the model prediction function that leads to errors due to backdoor poisoning attacks can be defined as:

$$F(wf,\theta_{po}): X \to Y, \text{ while } F(wf,\theta_{po}): X' \to Y' \tag{2}$$

where $\theta_{po}$ is the change in model parameters caused by training the model on the wrong dataset, such that when the model receives an input set $X$ that does not contain a trigger, it will output a set $Y$ of accurate predictions, as well as the original model without the wrong dataset. However, when the model receives an input set $X'$ containing triggers, it will output an incorrect prediction result set $Y'$, and the recognition accuracy will decrease significantly.

The model prediction function after adding the machine unlearning module can be defined as follows:

$$F(wf,\theta'): X \to Y, \text{ while } F(wf,\theta'): X' \to Y \tag{3}$$

where $\theta'$ is the final model parameter that is adjusted after the model after the machine unlearning module. Through machine unlearning, the model successfully forgets the influence of training with the wrong dataset. It can not only output the correct prediction result set $Y$ for the normal input set $X$, but also ensure the accurate prediction result $Y$ when receiving the input set $X'$ with triggers, which is in line with our research expectations.

## 3. Related Work

a) Website Fingerprinting attack

The inherent difficulty in discovering nodes, locating services, monitoring users, and confirming communication relationships makes dark web be widely used by criminals in cyber crimes and becomes a lawless digital space. Website Fingerprinting is an attack method against the hidden service mechanism of the dark web developed by network researchers. It can monitor the communication traffic between the client and the anonymous proxy (typically the onion router), which generally refers to the Tor unit used by onion routing. Attackers compare the traffic pattern generated in communication with the pre-collected traffic pattern of a specific website, and identify the website visited by the Tor user by analyzing the traffic characteristics, such as packet size and time information[20].

Website Fingerprinting is based on the premise that when different users visit the same website, the server always exhibits its static behavior[21][22], in other words, the same server generally does not generate different traffic packets for different visitors. Under this premise, packet sequences generated by different web servers can be considered as characteristics that distinguish them from other websites, including the traffic size, direction, length distribution, upstream/downstream byte patterns, burst traffic length, and so on[16][17]. Therefore, before launching an attack on a specific user, attackers often collect a large number of traffic instances from the target website and compose a training sample set after data processing. After feature extraction, these traffic samples are used to pre-train the WF model to obtain multiple classifiers that predict the labels of each target website. After collecting the traffic between the client and the anonymous proxy, the attacker will use this traffic as the input of a multiple classifier model, which outputs a predicted label for the visited website, as shown in Figure 2. Because the attacker does not take the initiative to destroy the information integrity and availability of the data packets, it remains highly covert and difficult for users to detect. However, in practice, the attacker can't collect the fingerprints for all the target websites, so they typically focus on fingerprinting and analyzing commonly visited sites likely to be accessed by their target users.

The first step in a Website Fingerprinting attack should be to determine whether the fingerprinting information is within the scope of the website monitored, that is, whether the attacker has pre-collected and trained the traffic characteristics associated with the website users are visiting. Therefore, when conducting relevant experimental research, the experimental environment is usually established under the following two assumptions: close-world assumption (CW) and open-world assumption (OW)[2][24][25]. In the closed-world assumption, the attackers will assume that the user will only visit websites where they have collected fingerprinting information in advance, and has also pre-trained the WF classifier associated with it, so the target website is always under his supervision when carrying out the attack. This assumption is often more suitable for verifying the effect of the designed model in the experiment more quickly and clearly. In the open-world assumption, the access request of the monitored user may exceed the website that the attacker can monitor. Blindly applying a CW-optimized model here can drastically reduce accuracy and potentially pollute the original model. Nevertheless, the OW assumption is closer to the real situation than the CW assumption and can evaluate the researchers' model realistically. This is also the experimental environment that Website Fingerprinting analysts must verify if they want to improve the robustness of the model. In our study, the CW and OW assumptions are both designed to evaluate the model more comprehensively.

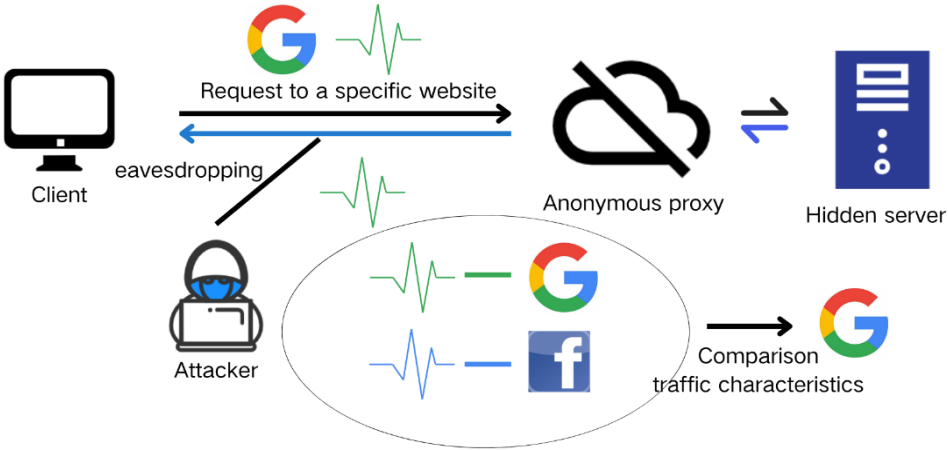

Figure 2 Illustration of Website Fingerprinting Attack

### b) Website Fingerprinting threat

Website Fingerprinting attack can effectively de-anonymize the dark web communication system like Tor[26], but it also faces many limitations in practical applications. On the one hand, non-human reasons such as network dynamic changes, network noise, and concept drift will lead to changes in network traffic characteristics, making the difference between the training samples and the actual test samples, affecting the accuracy and stability of the model. On the other hand, the hidden service mechanism of the dark web itself is designed to protect users' privacy and anonymization, making the traffic characteristics difficult to obtain and analyze. Most of the WF attacks can allow the collected data to have some noise, which is closer to the real network situation and shows the robustness of the model. In addition, the WF technology based on deep learning has achieved remarkable results in Walkie-Talkie defense, a special protection mechanism against Tor onion routing[3]. The Walkie-Talkie employs a half-duplex communication model merging the original traffic with the traffic of a randomly selected decoy page to mislead the WF analysis [Wang et al.[47]]. Although this defense will introduce a slight bandwidth and latency overhead, it can significantly enhance resistance to traditional WF feature extraction and is consequently widely adopted in Tor routing.

However, most of the WF attack models ignore the threat of actively dataset poisoning. State-of-the-art models, typically deep neural networks, incur high computational costs during training and demand substantial GPU resources, which makes many researchers choose to train on cloud servers. Some vandals of WF attacks can create a maliciously trained network (backdoor neural network)[36] and execute backdoor attacks on deep learning models. Models subjected to backdoors have state-of-the-art performance on standard training and validation samples from users but perform extremely poorly on inputs containing attacker-implanted trigger. Neural network backdoors are potent, and because the model behavior is difficult to explain, the victim is not easy to detect and eradicate. This type of backdoor poisoning attack is highly covert: it evades the standard model validation process and does not introduce any structural changes to the neural network trained on a clean dataset. However, once the input contains backdoor triggers, the model will necessarily produce an output that is quite different from the correct prediction [35][36].

Liang et al. [36] proposed adding backdoor triggers to the network traces that come from dark web communication (such as communication requests from clients to servers, traffic feedback from servers to clients, communication packets from Tor onion routing relay nodes, etc.). This is a novel way of poisoning, as each time the WF attacker collects data from compromised websites, the training dataset includes not only the clean website fingerprints and tags, but also some website sequences containing backdoor triggers. If the Website Fingerprinting model cannot correctly distinguish those website fingerprints injected with triggers, but trains them with the normal clean dataset, these seemingly insignificant poisoning data points will gradually affect the model parameters during the gradient training, making the final output of a poisoned model. Since the trigger has the following two characteristics[35][36]: 1. Triggers should be very easy to learn by the WF model and closely associated with the target website label. 2. Triggers are not easily detected and removed by WF analysts. Therefore, the poisoned model behaves like the normal model when receiving the WF input without the trigger, which is mainly due to the fact that the backdoor neural network has little impact on the original deep neural network architecture[35]. However, when we include the fingerprints of a website with a trigger as input, the classifier model performs very anomalously, predicting a drastically different result from the correct one even though the poisoned input is almost the same as the correct one.

### c) Unlearning

Machine unlearning refers to a series of methods and algorithms that enable a machine learning model to remove or forget specific information learned from particular subsets of their training data. At present, the mature application scenarios for machine unlearning mainly include: 1. Exposing the original data information; 2. Expose part of the information of whether the member participates in the training, namely the threat of member inference attack; 3. Model migration and traceability, i.e. when a machine learning model needs to be transferred from one environment to another, it may be necessary to forget the data from the previous environment to avoid data leakage or potential security issues.

Cao et al.[1] proposed the concept of unlearning in 2015. The original idea is to solve the privacy problem of AI models[4] [1]. The model security is guaranteed by forgetting and deleting the user's private information in the model training, thereby defending against attacks that destroy the confidentiality of information such as member inference attacks are defended, as shown in Figure 3. In 2024, Goel et al. [13] proposed the concept of "corrective forgetting", demonstrating the use of unlearning algorithms to address data errors in models caused by natural or human factors.

According to the forgetting effect of the model, machine unlearning is usually classified into two types in research: exact unlearning and approximate unlearning.

Exact unlearning refers to completely eliminating the influence of data on the model. The main idea is to isolate the data to be forgotten from the training process. This often involves complex modifications to ensure that the forgotten data does not have any impact on the model's predictions or analysis. To achieve exact unlearning, researchers usually need to improve the specialized algorithm of machine unlearning, or the process structure of algorithm operation, such as SISA (Shard Isolation Slice Aggregation model)[Sirinam et al. [4]], ACANE[Yan et al.[14]] and so on.

Approximate unlearning means that the influence of data on the model is approximately eliminated so that the unlearning model reduces the influence of the data to be forgotten below a certain threshold. This often involves fine-tuning the model parameters to minimize the impact of the forgotten data on the model. To achieve approximate unlearning, the parameters of the model are fine-tuned by analyzing the influence between training points, which includes the approximate unlearning based on the influence function. In addition, it can also act on the training process of the model, and machine unlearning based on gradient updating is performed on the basis of supervised model training gradients. Although this kind of idea does not reach the theoretical guarantees of exact unlearning, it can greatly save the unlearning time of the model and improve the operating efficiency, such as PUMA[Wu et al.[15]], Instance-wise unlearning[Cha et al.[16]], SSD[Foster et al.[19]], EU-k[Goel et al.[37]], CF-k[Goel et al.[37]], BadT [Kurmanji et al.[45]], SCRUB[Golatkar et al.[46]].

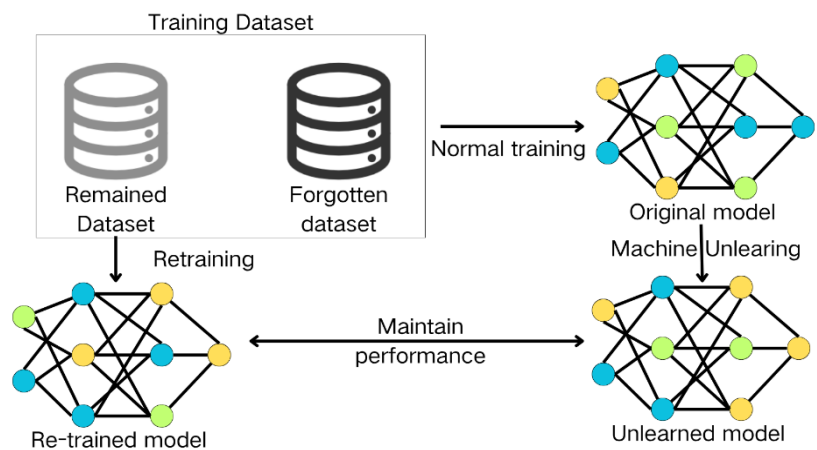

Fig 3 Illustration of Machine Unlearning

## 4. Method Design

a) Detect the poisoning spots

This section details the detection of the poisoned point module of the model. The current advanced WF attacks usually ignore the case that the training sample set used to train the classifier is injected by backdoor poisoning. This poisoning threat is often manifested in the results: the model exhibits normal, satisfactory accuracy on monitored websites when processing inputs without backdoor triggers; However, once there is a backdoor trigger in the monitored traffic, the performance of the WF classifier will be seriously abnormal, and the predicted value will significantly shift to the target website expected by the backdoor attackers, resulting in the paralysis of the model.

Because this kind of poisoning attack usually only adds a small amount of subtle filling to the training traffic of the Website Fingerprinting as a trigger content. Therefore, in the real situation, it takes a lot of time for the WF attacker to go through the training dataset once, while the effect of manual detection cannot meet the expectations, and the backdoor poisoning cannot be truly eliminated. Koh et al[49] proposed that a special influence function method can be used to understand the contribution value of training samples to the prediction output of the black-box model. This allows us to consider a more realistic scenario: If the WF trainer can find a small number of test data points that can be judged as poisoning (in the form of the test set that predicts the abnormal label), then we can calculate which training data points are most influential to these poisoned test samples, that is, which training data points are subject to the backdoor poisoning attack. Therefore, we use this calculated special contribution value as the influence score for the anti-identification of poisoned training data points. By using the abnormal test point set $(x_i, y_i) \in D_{ab} \subset D_{te}$ (the abnormal test point set $D_{ab}$ is a subset of the final total test set $D_{te}$ and $|D_{ab}| \ll |D_{te}|$), the forgotten dataset $D_{fo}$ is inverted to make $D_{fo}$ as close as possible to the poisoned dataset $D_{po}$.

To achieve accurate detection of backdoor poisoning points and pave the way for subsequent unlearning, we optimize the use of the influence score, using the initial quantification of the influence score for each training data point on the abnormal test output, and then screening the forgotten dataset according to the threshold. After calculating the first influence score, we perform network trace enhancement on the abnormal WF test points (mainly including inserting the same direction unit, splitting the different direction unit into the same direction, merging the same direction unit, reversing the direction of the unit and other unit operations for the website fingerprint, and randomly select several of them). At the same time, the enhanced fingerprints are randomly matched with the website labels in the test dataset. We can express such a transformation as $T = \{trans_i\}_{i=1}^{|D_{ab}|}$. Therefore, we reshape a new outlier test set $D'_{ab} = T(D_{ab})$ with such a transformation and then calculate the influence score of each training data point on the new outlier test set $D'_{ab}$ again. Since the clean WF points themselves undergo the operation of network trace enhancement after feature extraction, the change of the influence score is relatively stable. In contrast, for the poisoned data point, the embedded backdoor trigger cannot automatically adapt to the enhancement transformation of the test point. Consequently, the difference in their influence scores before and after augmentation becomes significantly amplified, which provides us with a signal to determine whether the data point is poisoned or not. This change can be expressed as follows:

$$diff = \frac{S((x_i, y_i)_{tr}, D_{ab})}{S((x_i, y_i)_{tr}, D'_{ab})} - 1 \qquad (4)$$

Compared with the direct use of influence score for threshold comparison, the optimized use of influence score expands the difference between clean data points and poisoned data points, and it will not bring a large burden to the model.

---
**Algorithm 1** Backdoor Poison Point Detection Algorithm
---
**Input:** the anomaly test set is known: d_ab, total training set of WF: d_tr, parameters of the WF model: p_po
**Output:** website fingerprint dataset to forget: d_fo
1:  **for** x, y in d_tr **do**
2:      **for** x, y in d_ab **do**
3:          score1 ← calculate influence score  # Compute the influence score before the transformation
4:          (x, y) ← trace augment random choose    # Applying network trace augmentation
5:              {insert, split, merge, flip}
6:          random match label y
7:          score2 ← calculate influence score    # Calculate the transformed influence score
8:          **if** diff > threshold
9:              d_fo ← (x, y)         # If it exceeds the threshold, it is judged as a poisoning point
10:         **end if**
11:     **end for**
12: **end for**
13: **return** d_fo
---

b) Calculate the parameter information contribution

Our unlearning method based on parameter tuning uses the Fisher Information Matrix (FIM) in the computation process. In mathematical theory, the Fisher Information Matrix can be regarded as the negative expected value of the Hessian Matrix of the model's log-likelihood function. The latter is generally used to optimize the objective function in the field of machine learning, which is represented by the second-order partial derivative matrix of the target model output concerning the model parameters. Therefore, FIM is often used to evaluate the sensitivity of model parameters.

Numerically, a larger Fisher information value for a parameter indicates that the parameter's estimation is more precise given the observed data, namely from the perspective of model parameters, the more important the parameter is for these model outputs. Therefore, the FIM can also be interpreted as the contribution of each parameter for the specific output prediction. As shown by Kirkpatrick et al.[42]: the FIM is used to calculate the regularization term to prevent the model from forgetting what it has previously learned.

Golatkar et al. [48] pioneer the use of FIM for unlearning by proposing noise injection proportional to the FIM's diagonal elements computed over both retained and forgotten datasets (specifically, injecting small noise or no noise into the parameters that are strongly correlated with the retained dataset and weakly correlated with the forgotten dataset). The noise injected into the model parameters will induce the model to forget the content of the non-reserved dataset that has been trained and learned during model verification and testing so that the model cannot respond and give feedback to this part of the data. This kind of method has a large amount of calculation in the specific implementation, and it is expensive in time for needing to go through every global model parameter and coordinate the update. At the same time, when measuring those models that are important to both the retained dataset and the forgotten dataset, or are not important to the two datasets, it cames errors, result in affecting the recognition accuracy of the model on the retained dataset.

We use the Fisher Information Matrix in a way that does not need to modify the global model parameters, which significantly accelerates the execution efficiency of the unlearning algorithm. In addition, it does not need to rely on other additional models, which simplifies the complexity of the overall unlearning algorithm and facilitates application deployment.

We summarize the mathematical expression of the FIM: given a model parameter $\theta \in \mathbb{R}^K$, $K$ is the total number of parameters. According to the above theory, the contribution value of the model parameters to the data points on $D_{tr}$ can be described by the Fisher information, which is expressed as the estimation accuracy of $D_{tr}$ on $\theta$.

c) Parameter-based unlearning

Feldman et al.[43]、Stephenson et al.[44] have shown that when deep neural networks memorize specific training examples, parameters in later network layers specialize to fit these distinct features. As a result, the fitted parameters may be important for a small subset of the training dataset, but their importance decreases significantly for a larger subset of the training dataset. As to the backdoor poisoning injection attack scenario on WF, the whole training sample set $D_{tr}$ is usually large and contains a large amount of data, while the forgotten dataset $D_{fo}$ composed of real poisoning data points has more specific characteristics compared with other data points. According to the above theory, the final poisoned WF classifier model must contain some specialized parameters, that show a high degree of importance to the poisoned data points, and it will not pose an obvious impact or threat to the learning of the data points in the reserved dataset $D_{re}$ with more universal and highly generalized features. For example, in the poisoning scenario of WF, ordinary website fingerprints corresponding to various websites such as Google, Facebook, YouTube, etc. have many similarities, but those website fingerprints injected with backdoor triggers will cause the model to specialized memory learning.

In the part of Top-K parameter selection, the core idea of instructing the model to operate is to find the Top-K specialized model parameters that are not very important for the model to retain the dataset $D_{re}$, but are highly important for the model to forget the dataset $D_{fo}$. K can be set according to user requirements. According to the expression of the Fisher Information Matrix, we can obtain the information contribution value description of $D_{re}$ and $D_{fo}$ to the exposure model parameter $\theta_{po}$.

Therefore, we use parametry-based unlearning technology to forget the detected poison spots, and the purpose is to suppress the Top-K specialization parameters that contribute greatly to $D_{fo}$ but less to $D_{re}$ in proportion to the contribution value gap. Because in our backdoor poisoning attack unlearning scenario, the number of poisoning points infected by malicious attackers is very small (backdoor injectors generally only select a small part of the training dataset for backdoor trigger injection, to reduce the possibility of being discovered by the model owner) [44]. Therefore, the worst case is merely the wrong suppression of this model parameter, which makes the performance of the WF analysis model relatively unstable under the clean dataset, but it can still have a satisfactory detoxification success rate and ensure the accuracy of WF attack on the poisoned dataset. In addition, whether it is the WF training dataset $D_{tr}$, the forgetting dataset $D_{fo}$, or the retained dataset $D_{re}$, in the process of model execution, it usually need to be calculated once or twice, without additional storage. On the one hand, it improves the space utilization of the WF attack model and greatly reduces the time cost. On the other hand, it reduces the risk of data leakage when stored locally or in the cloud.

---
**Algorithm 2** Parameter-based unlearning algorithms
---
**Input:** WF dataset needed to maintain performance: d_re, WF dataset to forget: d_fo, total training set of WF: d_tr, set of parameters of the WF analysis model: p_po
**Output:** the final set of model parameters after performing unlearning: p_unlearning

```
 1:  for p in p_po do
 2:      for x, y in d_re do
 3:          imp_re ← importance calculate     # The information contribution of the retained dataset
 4:      end for
 5:      for x, y in d_fo do
 6:          imp_fo ← importance calculate     # The information contribution of the forgotten dataset
 7:      end for
 8:      if imp_fo > imp_re and Top-K is satisfied then
 9:          p_mo ← p                          # Choose the Top-K parameters
10:      end if
11:  end for
12:  for p in p_mo do
13:      for x, y in d_tr do
14:          imp_tr ← importance calculate     # The information contribution for the full training dataset
15:          P ← update
16:          p_unlearning ← p
17:      end for
18:  end for
19:  return p_unlearning
```

# 5. Experiments

a) Experimental Setup

　　i.　Website Fingerprinting dataset

In order to verify the effectiveness of this research scheme, we selected tor_100w_2500tr and tor_open_200w_2000tr datasets used in the WFlib Website Fingerprinting attack[6], which correspond to two assumptions in WF scenarios: close-world assumption and the open-world assumption, which are defined in 3.1 Website Fingerprinting attack. The CW dataset tor_100w_2500tr contains about 250,000 fingerprints of dark web websites, involving 100 different website labels. The OW dataset tor_open_200w_2000tr contains about 400,000 fingerprints of dark web websites, involving 200 different website labels, and contains unknown

WF samples. The input of these datasets is in the form of Tor unit sequences of onion routing, which mainly contains the number and direction characteristics of traffic, and the output is the website labels corresponding to these website fingerprints, which are obtained from the browsing records of popular dark web users. Our experiments will be conducted on both types of datasets.

Under the backdoor poisoning attack, the original WF dataset is divided into two parts, a small part of which is the WF samples that are successfully injected into the backdoor trigger, and the majority of which are completely clean. This experiment will first use the mixed dataset to train the WF analysis model, and then carry out verification tests on the divided poisoned dataset and clean dataset, as well as the test set mixed with backdoor triggers and clean samples, so as to prove the feasibility of our model.

ii. Methods of comparison

On the overall WF analysis model, we choose three WF attack algorithms, DF, TF, and AWF, as the comparison model algorithm.

DF（Deep Fingerprinting）[3]: The algorithm mainly adopts deep learning mode and uses a convolutional neural network to train the WF classifier model, which has high accuracy in the general closed-world assumption and the open-world assumption.

TF（Triplet Fingerprinting）[5]: The algorithm used the triplet network to carry out the WF attack, and by designing positive and negative examples, it could deal with the situation of less labeled training samples. In the process of feature extraction, the model can combine with the KNN model to generate embeddings for subsequent training.

AWF（Automated Website Fingerprinting）[6]: The meta-features of WF are automatically extracted by deep learning algorithms, and trained by models such as convolutional neural network (CNN) and long short-term memory network (LSTM). By hierarchically learning the abstract patterns in traffic, the unique fingerprintings of different websites are automatically identified, to realize the accurate classification of users' visits to websites.

These three WF attack algorithms are the mainstream representatives of integrating deep learning technology into WF and can be compared with this research scheme to verify their effectiveness and robustness.

iii. Metrics

In this experimental scheme, the following four key indicators are selected to evaluate the performance of the WF analysis model:

1. Accuracy on the taining dataset (on the dataset injected with the backdoor trigger): expressed as the ratio of the number of samples that were correctly identified as the original website label (rather than the new website label pointed to by the backdoor trigger) to the number of samples injected with the backdoor trigger.

2. Accuracy on the clean dataset: expressed as the ratio of the number of samples that were correctly identified as the label of the corresponding website to the number of all samples that were not injected with backdoor triggers.

3. Test accuracy: Manifested in the test set with possible backdoor triggers. 1. If it is a sample that is injected into the backdoor trigger, count the number of samples that are correctly identified as the original website label; 2. If no backdoor trigger is injected, then count the number of samples that are accurately identified as the label of the corresponding website. The value of the test accuracy is equal to the ratio of the sum of the number of samples in the case of statistics one and two and the test samples of all website fingerprintings.

4. Execution time: For the comparison experiment of the overall framework, after statistical data preprocessing (backdoor injection has been completed), from the model pre-training (i.e., feature extraction of some WF analysis models, pre-training of unlearning, and data augmentation of some WF analysis models), to the output of the final results at the end of the model operation, The operating efficiency of each algorithm model was compared.

Each experimental data is the average obtained after three tests with different random seeds.

b) Analysis of experimental results under the CW

In this experiment, we use the dataset tor_100w_2500tr to show the experimental results of the model under the closed-world assumption (the website labels appearing in the test set and validation set must exist in the training set). Our model framework is compared with the three WF attack frameworks mentioned in the experimental setup in the context of backdoor poisoning attacks to verify the effectiveness of our WF attack against backdoor poisoning.

Table 1 shows the performance of various WF attack models in a complex and suboptimal scenario that includes both clean data and backdoor poisoning injection attacks (3000 poisoned data points and 300 known poisoning test points). Figure 4 shows the performance comparison of various WF attack frameworks in the CW scenario more intuitively.

Table 1: Performance comparison of WF attack model under CW in backdoor poisoning scenario (accuracy in percentage)

| Model method | Poisoned dataset accuracy | Clean dataset accuracy | Test accuracy | Time/s |
| --- | --- | --- | --- | --- |
| DF | 3.84 | 99.34 | 12.4 | 715.23 |
| TF | 1.86 | 99.15 | 0.52 | 868.43 |
| AWF | 8.02 | 96.56 | 44.72 | 678.24 |
| **Ours** | **87.03** | **96.26** | **83.24** | **297.53** |

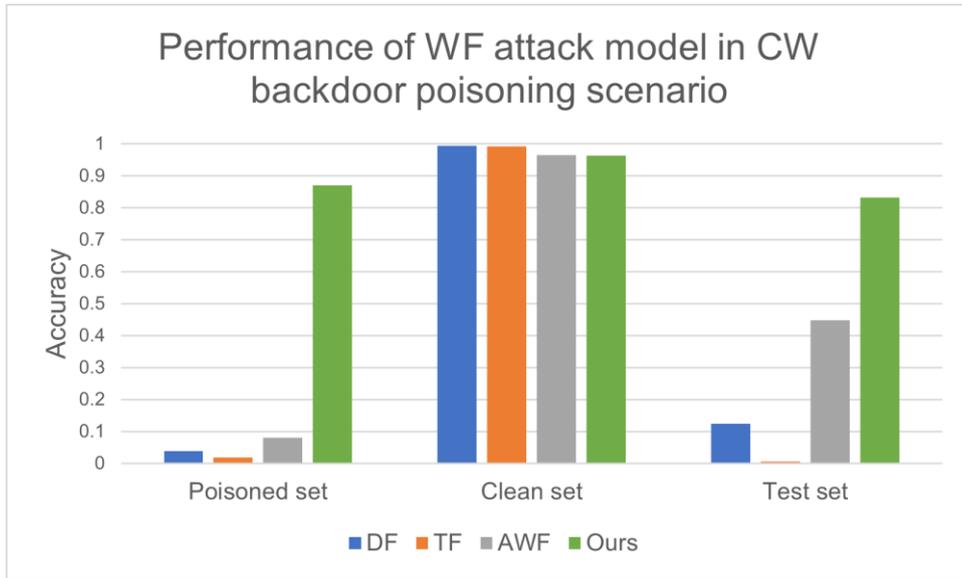

Figure 4 Performance of WF attack model in CW backdoor poisoning scenario

As can be seen from Figure 4, when compared with other WF attack models, our method can effectively remove the poisoning effect caused by backdoor triggers under the premise of maintaining a high recognition rate of clean datasets, which is not done by other classifier models based on deep learning. Combined with Table 1, the WF attack experiment is also carried out in an unsatisfactory environment. The method used in this paper not only achieves remarkable results in the detoxification level but also achieves better execution speed while resisting backdoor poisoning.

c) Analysis of experimental results under the OW

Similarly to the experimental content and indicators of the closed-world assumption, the dataset tor_open_200w_2000tr will be used to show the experimental results of the model in the scenario of the open-world assumption (the website labels in the test set and the validation set may not exist in the training set, that is, there are unknown and unmonitored website fingerprintings).

Table 2 shows the performance of various WF attack models in a complex and suboptimal scenario under OW that contains both clean data and backdoor poisoning injection attacks (3000 poisoning data points in total and 300 known poisoning test points). Figure 5 shows the performance comparison of various WF attack frameworks under OW more intuitively.

Table 2: Performance comparison of WF attack model under OW in backdoor poisoning scenario (accuracy in percentage)

| Model method | Poisoned dataset accuracy | Clean dataset accuracy | Test accuracy | Time/s |
| --- | --- | --- | --- | --- |
| DF | 0.94 | 98.87 | 4.10 | 1143.45 |
| TF | 0.54 | 98.02 | 0.78 | 1447.77 |
| AWF | 4.84 | 96.06 | 39.22 | 1054.00 |
| **Ours** | **83.80** | **94.82** | **79.48** | **434.47** |

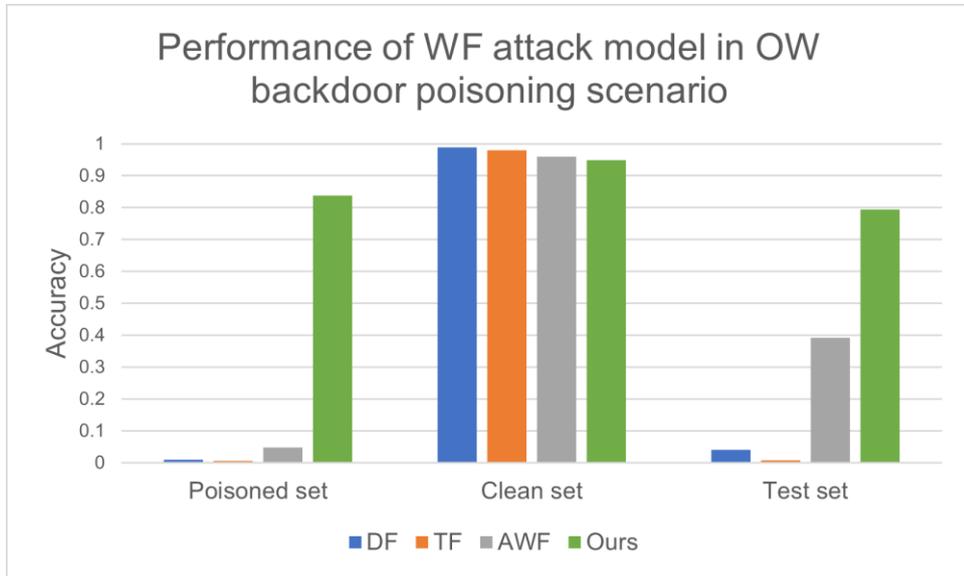

Figure 5 Performance of WF attack model in OW backdoor poisoning scenario

Combined with Table 2 and Figure 5, it can be seen that under the premise of the open-world assumption, the performance of the WF attack model has suffered certain damage compared with the closed-world assumption, especially the accuracy of the poisoned dataset and the test set has decreased by nearly 5%. Nevertheless, our method can stabilize the horizontal line with 80% accuracy, while the other WF analysis models, except AWF, perform very poorly on the test set, with an accuracy close to 0%. Although AWF can achieve close to 40% test accuracy, it cannot cope with the corruption of taining data points. In addition, due to the larger dataset of the open-world assumption and the more complex situation (there are unknown websites), the execution time of all models increases by different degrees. Among them, our execution time increases the least, and the final execution speed is about twice that of all models, which ensures high operating efficiency while maintaining a high recognition rate.

d) Ablation experiment

In order to verify that our method indeed plays a significant detox effect on the improvement of the poisoning model, this experiment conducts ablation experiments on two key structures: the parameter selection part and the parameter-based unlearning part. Because our method itself can refine the trade-off between model forgetting effect, model retention performance, and execution speed, it is not necessary to re-change the code and structure of the model when performing the ablation of specific modules.

If you leave the parameter selection part out, the model will indiscriminately include almost all of the model parameters in the suppression list, which is equivalent to throwing out the effect of the parameter selection part.

To discard the parameter-based unlearning part, we simply zero out all the selected parameters, that is, instead of doing a proper suppression reduction on the target parameter, we discard the parameter completely.

Table 3 and Table 4 respectively show the results of ablation experiments under the closed-world assumption and the open-world assumption when the number of poisoned data points is set to 2500, highlighting the importance of the core part of this research protocol.

Table 3: Ablation experiment records under CW (accuracy in percentage)

| Model method | Poisoned dataset accuracy | Clean dataset accuracy | Test accuracy | Time/s |
| --- | --- | --- | --- | --- |
| Without selection | 72.36 | 90.01 | 70.12 | 10.69 |
| Without inhibition | 82.32 | 94.28 | 79.24 | 10.70 |
| Baseline | 86.83 | 95.85 | 83.02 | 10.60 |

Table 4: Ablation experiment records under OW (accuracy in percentage)

| Model method | Poisoned dataset accuracy | Clean dataset accuracy | Test accuracy | Time/s |
|---|---|---|---|---|
| Without selection | 26.24 | 69.48 | 26.17 | 18.03 |
| Without inhibition | 67.12 | 91.73 | 65.61 | 18.30 |
| Baseline | 83.88 | 94.98 | 79.93 | 18.42 |

Combined with Table 3 and Table 4, firstly, it can be seen that neither parameter selection nor parameter suppression have a significant impact on the execution time, indicating that these two parts will not significantly affect the operation efficiency.

When the model parameters are not restricted, the accuracy of the model decreases greatly in both CW and OW scenarios. Especially in OW, the model is reduced from 83.88% and 79.93% to about 26%, which even affects the recognition accuracy of the clean dataset, reducing it by about 20%. Therefore, it can be considered that parameter selection is very critical for the model in this study.

If the selected model parameters are directly zeroed out, the experimental data show that it mainly affects the accuracy of the poisoned dataset and the test accuracy, and has little impact on the clean dataset. Therefore, it is believed that the parameter modification part is less important than the parameter selection part, and it mainly affects the detoxification function of the model, and has a small proportion of the influence on the retention function.

e) Experimental conclusions

Our WF attack model, integrated with unlearning technology, achieves a significant backdoor poisoning removal effect in both closed-world assumption (Section 5.2 experiment) and open-world assumption (Section 5.3 experiment), and maintains the performance of the model itself on clean datasets. For the complex and unsatisfactory scenes mixed with clean WF features and backdoor triggers, the test performance of our model can still achieve a satisfactory state while other comparison methods suffer a significant decrease in accuracy, which fully proves its feasibility and robustness. In addition, through the comparison of execution speed, whether it is a separate test of machine unlearning part or the overall framework test of WF analysis, this research scheme shows significant advantages in execution efficiency. Finally, in the ablation experiment in Section 5.4, using the control variable method, the necessity of incorporating parameter-based unlearning is demonstrated.

# 6. Conclusion

To address the critical vulnerability of existing Website Fingerprinting attack techniques against poisoning attacks in dark web anonymous communication systems, we propose an optimized framework integrating machine unlearning for automatic poison detection and detoxification. This approach effectively mitigates the threat of backdoor poisoning attacks to model robustness. Our key contributions are:

**1. Apply unlearning technology in the field of WF:** experiments show that with the effective integration of WF attack and unlearning, our model significantly outperforms traditional WF attack schemes for backdoor poisoning data points under both closed-world (CW) and open-world (OW) assumptions. When the accuracy of the poisoned dataset of other methods is lower than 50% or even close to 0%, our classification accuracy is stable at over 80% respectively, achieving satisfactory results.

**2. Few-shot poisoned point detection method:** in the experiment, our optimized detection method integrated with unlearning realizes the elimination of most backdoor poisoning data points on the premise of fewer known poisoned test points, and does not significantly increase the execution time of the model. It is verified that our model has higher detection efficiency of backdoor poisoning and is closer to the actual needs.

**3. Higher backdoor poisoning and forgetting capabilities:** compared with other WF attack methods, we achieve a better-forgetting effect (consistently better than existing methods) for different numbers of poisoned points in both closed-world and open-world, and can maintain clean dataset performance (about 95% recognition accuracy).

**4. More efficient WF model:** experiments show that our scheme runs 2-3 times faster than other methods in both closed-world and open-world settings.

Combined with the current research work, the model implemented in this paper still has some shortcomings, and future research can be further explored from these scenes: 1. The generality of the machine unlearning algorithm should be optimized. The current model was only suitable for backdoor poisoning attacks, and the integration with differential privacy and other technologies could be explored in the future to deal with more complex adversarial attack scenarios. 2. Our scheme needs to further improve the robustness of the model under unknown network fluctuations and natural updates of website content. 3. We also need to pay attention to the engineering challenges in practical deployment, such as adaptive hyperparameter adjustment, distributed computing efficiency under large-scale data, etc.

With the evolution of dark web anonymity technology, it is hoped that the technical framework of this research can provide theoretical support for cyberspace security governance and generate wider social value in cross-domain applications such as data detoxification and privacy protection.

# References

[1] Joel John, Chainalysis: 2022 Cryptocurrency Survey Report [R]. America, Chainalysis.

[2] Dyer K P, Coull S E, Ristenpart T, et al. Peek-a-Boo, I still see you: Why efficient traffic analysis countermeasures fail[C] Proc of the 33rd IEEE Symp on Security and Privacy. Piscataway, NJ: IEEE,2012: 332-346.

[3] P. Sirinam, M. Imani, M. Juarez, and M. Wright. Deep fingerprinting: Undermining website fingerprinting defenses with deep learning[C]. ACM CCS, 2018.

[4] Lucas Bourtoule, Varun Chandrasekaran, Christopher A. Choquette-Choo, Hengrui Jia, Adelin Travers, Baiwu Zhang, David Lie, Nicolas Papernot. Machine Unlearning[C]. IEEE Symposium on Security and Privacy (SP), 2021.

[5] P. Sirinam, N. Mathews, M. Rahman, and M. Wright. Triplet Fingerprinting: More Practical and Portable Website Fingerprinting with N-shot Learning[C]. In ACM CCS, 2019.

[6] Vera Rimmer, Davy Preuveneers, Marc Juarez, Tom Van Goethem and Wouter Joosen. Automated Website Fingerprinting through Deep Learning[C]. Network and Distributed System Security (NDSS), 2018.

[7] Alireza Bahramali, Ardavan Bozorgi, Amir Houmansadr. Realistic Website Fingerprinting By Augmenting Network Traces[C]. ACM Conference on Computer and Communications Security, 2023, 11: 1035-1049.

[8] Kirkpatrick, J.; Pascanu, R.; Rabinowitz, N.; Veness, J.; Desjardins, G.; Rusu, A. A.; Milan, K.; Quan, J.; Ramalho, T.; Grabska-Barwinska, A.; et al. 2017. Overcoming catastrophic forgetting in neural networks[C]. Proceedings of the national academy of sciences, 114(13): 3521–3526.

[9] Maymounkov P, Mazieres D, Kademlia, A peer-to-peer information system based on the metric [C]. Proc of the 1st Int Workshop on Peer-to-Peer Systems. Berlin: Springer, 2002: 53-65.

[10] McLachlan J, Tran A, Hopper N, et al. Scalable onion routing with Torsk[C]. Proc of the 16th ACM Conf on Computer and Communications Security. New York: ACM, 2009: 590-599.

[11] Dingledine R, Mathewson N, Syverson P. Tor: The second-generation onion router[C]. Proc of the 13th USENIX Security Symp. Berkeley, CA: USENIX Association, 2004: 1-18.

[12] Danezis G, Dingledine R, Mathewson N, Mixminion: Design of a type III anonymous remailer protocol[C]. Proc of the 24th IEEE Symp on Security and Privacy. Piscataway, NJ: IEEE, 2003: 2-15.

[13] Shashwat Goel, Ameya Prabhu, Philip Torr, Ponnurangam Kumaraguru, Amartya Sanyal. Corrective Machine Unlearning[C]. ICLR, 2024.

[14] Haonan Yan, Xiaoguang Li, Ziyao Guo, Hui Li, Fenghua Li, Xiaodong Lin. ARCANE: An Efficient Architecture for Exact Machine Unlearning[C]. International Joint Conference on Artificial Intelligence, 2022: 4006-4013.

[15] Ga Wu, Masoud Hashemi, Christopher Srinivasa. PUMA: Performance Unchanged Model Augmentation for Training Data Removal[C]. AAAI Conference on Artificial Intelligence, 2022: 8675-8682.

[16] Sungmin Cha, Sungjun Cho, Dasol Hwang, Honglak Lee, Taesup Moon and Moontae Lee. Learning to Unlearn: Instance-wise Unlearning for Pre-trained Classifiers[C]. AAAI Conference on Artificial Intelligence, 2024.

[17] Ishan Karunanayake, Jiaojiao Jiang, Nadeem Ahmed, Sanjay K. Jha. Exploring Uncharted Waters of Website Fingerprinting[J]. IEEE Journals & Magazine, 2023.12: 1840 - 1854.

[18] Youngsik Yoon, Jinhwan Nam, Hyojeong Yun, Jaeho Lee, Dongwoo Kim, Jungseul Ok. Few-shot Unlearning[C]. IEEE Symposium on Security and Privacy, 2024.5.

[19] Jack Foster, Stefan Schoepf, Alexandra Brintrup. Fast Machine Unlearning Without Retraining Through Selective Synaptic Dampening[C]. AAAI Conference on Artificial Intelligence, 2024.

[20] Avarikioti Z, Pietrzak K, Salem I, Schmid S, Tiwari S, Yeo M. Hide & seek: Privacy-preserving rebalancing on payment channel networks[J]. Springer; 2022:358-373.

[21] Yuwen Qian, Guodong Huang, Chuan Ma, Member, Ming Ding, Senior Member, Long Yuan, Zi Chen, Kai Wang. Enhancing Resilience in Website Fingerprinting: Novel Adversary Strategies for Noisy Traffic Environments[J]. IEEE Transactions on Information Forensics and Security, 2024, 7: 7216 - 7231.

[22] P. Sirinam, M. Imani, M. Juarez, and M. Wright. Deep fingerprinting: Undermining website fingerprinting defenses with deep learning[C]. ACM CCS, 2018.

[23] S. Bhat, D. Lu, A. Kwon, and S. Devadas. Var-CNN and DynaFlow: Improved Attacks and Defenses for Website Fingerprinting[C]. arXiv preprint arXiv:1802.10215, 2018.

[24] V. Rimmer, D. Preuveneers, M. Juarez, T. Van, and W. Joosen. Automated website fingerprinting through deep learning[C]. NDSS, 2018.

[25] R. Dingledine, N. Mathewson, and P. Syverson. Tor: The second-generation onion router[C]. USENIX Security, 2004.

[26] J. Hayes and G. Danezis. k-fingerprinting: A robust scalable website fingerprinting technique[C]. USENIX Security, 2016.

[27] S. Ioffe and C. Szegedy. Batch normalization: Accelerating deep network training by reducing internal covariate shift[C]. ICML, 2015.

[28] M. Juarez, S. Afroz, G. Acar, C. Diaz, and R. Greenstadt. A critical evaluation of website fingerprinting attacks[C]. ACM CCS, 2014.

[29] S. Laine and T. Aila. Temporal ensembling for semi-supervised learning[C]. arXiv preprint arXiv: 1610.02242, 2016.

[30] D Lee. 2013. Pseudo-label: The simple and efficient semi-supervised learning method for deep neural networks. In ICML 2013 Workshop: Challenges in Representation Learning[C] WREPL, 2013.

[31] M. Sajjadi, M. Javanmardi, and T. Tasdizen. Regularization with stochastic transformations and perturbations for deep semi-supervised learning[C]. Advances in neural information processing systems, 2016.

[32] F. Schroff, D. Kalenichenko, and J. Philbin. Facenet: A unified embedding for face recognition and clustering[C]. CVPR, 2015.

[33] H. Scudder. Probability of error of some adaptive pattern-recognition machines. IEEE Transactions on Information Theory[C], 11, 3, 363–371. doi: 10.1109/TIT.1965.1053799, 1965.

[34] P. Sirinam, M. Imani, M. Juarez, and M. Wright. Deep fingerprinting: Undermining website finge

rprinting defenses with deep learning[C]. ACM CCS, 2018.

[35] Tianyu Gu, Brendan Dolan-Gavitt, Siddharth Garg. BadNets: Identifying Vulnerabilities in the Machine Learning Model Supply Chain[C]. Cryptography and Security, 2017.

[36] Siyuan Liang, Jiajun Gong, Tianmeng Fang, Aishan Liu, Tao Wang, Xianglong Liu, Xiaochun Cao, Dacheng Tao, Chang Ee-Chien. Red Pill and Blue Pill: Controllable Website Fingerprinting Defense via Dynamic Backdoor Learning[C]. Cryptography and Security, 2024.

[37] Shashwat Goel, Ameya Prabhu, Amartya Sanyal, Ser-Nam Lim, Philip Torr, Ponnurangam Kumaraguru. Towards Adversarial Evaluations for Inexact Machine Unlearning[C]. arXiv:2201.06640, 2022.

[38] Xinhao Deng, Qi Li, Ke Xu. Robust and Reliable Early-Stage Website Fingerprinting Attacks via Spatial-Temporal Distribution Analysis[C]. CCS, 2024.

[39] Nhien Rust-Nguyen, Mark Stamp. Darknet Traffic Classification and Adversarial Attacks[C]. arXiv:2206.06371, 2022.

[40] Lee, J.; Lee, J.-N.; and Shin, H. The long tail or theshort tail: The category-specific impact of eWOM on sales distribution[J]. Decision Support Systems, 2011, 51(3): 466–479.

[41] Maltoni, D.; and Lomonaco, V. Continuous learning in single-incremental-task scenarios[J]. Neural Networks, 2019, 116: 56–73.

[42] Kirkpatrick, J.; Pascanu, R.; Rabinowitz, N.; Veness, J.; Desjardins, G.; Rusu, A. A.; Milan, K.; Quan, J.; Ramalho, T. Grabska-Barwinska, A. Overcoming catastrophic forgetting in neural networks. Proceedings of the national academy of sciences[J], 2017, 114(13): 3521–3526.

[43] Feldman, V. Does learning require memorization a short tale about a long tail[C]. Proceedings of the 52nd Annual ACM SIGACT Symposium on Theory of Computing, 2020, 954–959.

[44] Stephenson, C.; Padhy, S.; Ganesh, A.; Hui, Y.; Tang, H.; and Chung, S. On the geometry of generalization and memorization in deep neural networks[C]. arXiv preprint,2021, arXiv:2105.14602.

[45] Meghdad Kurmanji, Peter Triantafillou, and Eleni Triantafillou. Towards unbounded machine unlearning. Advances in Neural Information Processing Systems (NeurIPS), 2023. 2, 3, 5, 9, 10

[46] Aditya Golatkar, Alessandro Achille, and Stefano Soatto. Forgetting outside the box: Scrubbing deep networks of information accessible from input-output observations. In European Conference on Computer Vision, 2020. 5, 10

[47] WANG, T., AND GOLDBERG, I. Walkie-talkie: An efficient defense against passive website fingerprinting attacks. In USENIX Security Symposium (2017), USENIX Association, pp. 1375–1390.

[48] Golatkar, A.; Achille, A.; and Soatto, S. Eternal sunshine of the spotless net: Selective forgetting in deep networks. In Proceedings of the IEEE/CVF Conference on Computer Vision and Pattern Recognition, 2023, 9304–9312.

[49] Pang Wei Koh, Percy Liang. Understanding Black-box Predictions via Influence Functions. In Proceedings of the 34th International Conference on Machine Learning, 2017b. 2, 3, 16

[50] Yinzhi Cao and Junfeng Yang. Towards making systems forget with machine unlearning. In IEEE Symposium on Security and Privacy (IEEE S&P), 2015. 16